\documentclass[10pt]{article}
\usepackage{amsfonts,amsmath,amssymb,graphics,epsfig}
\usepackage{enumerate}
\usepackage{theorem}
\numberwithin{equation}{section}

\begin{document}
\title{ {\large {\bf Friedman-like Robertson-Walker model in generalized metric space-time with weak anisotropy.}}   }
\author{{\large  P.C.Stavrinos }\footnote{pstavrin@math.uoa.gr},\\
{\small \emph{Department of Mathematics University of Athens 15784 Greece}} \\
{\large  A.P.Kouretsis}\footnote{a.kouretsis@yahoo.gr},\\
{\small \emph{ 5 Festou St. 16674 Glifada Athens  Greece}} \\
{\large
M.Stathakopoulos}\footnote{michaelshw@yahoo.co.uk},\\
{\small \emph{1 Anastasiou Genadiou St. 11474 Athens Greece}}}
\date{}
\maketitle
\begin{abstract}
\noindent {\small A generalized FRW model of space-time is studied,
taking into consideration the anisotropic structure of fields
which are depended on the position and the direction (velocity).The Raychaudhouri and
Friedman-like equations are investigated assuming the Finslerian character of space-time.A
long range vector field of cosmological origin is considered in relation to a physical
geometry where the Cartan connection has a fundamental role.The
Friedman equations are produced including extra anisotropic terms.The
variation of anisotropy $z_t$ is expressed in terms of the Cartan
torsion tensor of the Finslerian manifold.A physical generalization of
the Hubble and other cosmological parameters arises as a direct consequence of the equations
of motion. \newline\newline {\bf Key
words}:Finsler Geometry,Cosmology,Gravitation}\newpage
\end{abstract}

\section{Introduction}
During the last few years considerable studies concerning
observable anisotropies of the universe have been
investigated\cite{Wmap,MV,newref1,newref2}.These are connected to the very early
state of the universe and related to the estimations of WMAP of
CMB, the anisotropic pressure or the incorporation of a primordial
vector field(e.g. a magnetic field) to the metrical spatial
structure of the universe\cite{tsagas2,tsagasmaart,Maartens}.In this case the form of scale
factor can be influenced by the introductory field. A geometry
which may connect the Riemannian metric structure of the
space-time to physical vector fields, is the class of
Finsler-Randers type spaces. In these spaces an electromagnetic
field, a magnetic field or a gauge vector field may emerge out
by a physical source of the universe and can be incorporated into
the geometry causing an anisotropic
structure\cite{Prezas,StavJPh,Stavcong,cliff,phoevos,beil,Asanov}.

Finsler geometry or the theory of Finsler spaces may be considered
as a generalized Riemannian geometry of the first order within the
sphere of metrical differential geometry\cite{RundFB}.A Finsler space is a
metric space in which the metric function is defined by a norm
$\mathcal{F}$ on a tangent bundle instead of defining an inner
product structure on it. The norm will be a real function
$\mathcal{F}(x,y)$ of a space-time point $x$ and a tangent
vector $y \in T_x M$ which places the role of an internal
variable (Appendix A). This $y$ dependence characterizes
essentially the Finslerian field and has been combined with the
concept of anisotropy which causes the deviation from Remannian
geometry\cite{Horv1,Horv2}.All kinds of generalized metric theories and
unified field theories belong to the larger class of the so called
{\it anisotropic} field theories\cite{Horv1,Horv2}.Therefore these
geometrical anisotropies are caused by internal variables. Under
these conditions a Finsler geometry can be considered as a physical
geometry on which matter dynamics takes place while the Riemann
geometry is the gravitational geometry\cite{vacannals, cliff, bekenstein} .

The Cartan's torsion tensor\cite{RundFB,Asanov} characterizes all the
geometrical concept of Finsler geometry and appears to all
expressions of  geometrical objects such as connection and
curvature. In some cases it is useful from a physical point of view
consider a vector field in the form $y^i(x)/i=1,2,3,4$ and the
induced Finslerian metric tensor gives rise to the osculating
Riemannian metric tensor $r_{\mu\nu}(x)=f_{\mu\nu}(x,y(x))$.

In the present paper we adopt such an approach in order to obtain
some results concerning to a Friedman-like Robertson Walker
cosmological model(F-LRW) and its consequences. We
proceed by introducing a Randers-type Lagrangian and the induced
Finslerian metric modified appropriately for a weak primordial
vector field $u_a$\cite{phoevos} (sec.2). We construct the field
equations using the osculating approach\cite{RundFB,Asanov} and derive a
Friedman-like equation of motion with an extra anisotropic term
(sec.3,4). We generalize the Hubble parameter H, the density
parameter $\Omega$ and the deceleration parameter $q$ for a weak
anisotropic approach. The cosmological parameters depend directly
on the anisotropy generated by the vector field defined in the
Lagrangian metric function (sec.4). The solutions of the
Friedman-like equation for both matter and radiation dominated
universe, the Raychaudhuri equation initially presented in
\cite{StavJPh,Stavcong} and the CMB temperature estimation are all affected by
the presence of the rate of anisotropy at the field equations
(sec.4,5). The anisotropic solution of the scale factor coincides
with the standard ones derived under the assumptions of homogeneity and
isotropy. A possible estimation of the rate of
anisotropy parameter might be possible if we consider intermediate values
for the cosmological constant for a generalized de-Sitter model
defined by the Friedman-like equation of motion with a
cosmological constant\cite{diInv}.A more accurate estimation of the model's extra parameter
can be made by calculating the CMB shift and the baryon oscillation acoustic peak using
the Friedman equations and comparing them to the corresponding values coming from the WMAP
data\cite{Fair_Goob,Eisenstein,Wang,Mart_MG,tywmap}. The data produced by these tests depicts a
flat universe confirming the observations.

\section{Anisotropy and a Randers type Lagrangian metric}

An alternative way of studying of physical phenomena is to
incorporate the dynamics to the active geometrical background
following  Einstein's meaning of gravity. Our
investigation is based on the introduction of a Lagrangian
metric\cite{Stavcong,phoevos} considering the anisotropy of the universe\cite{Wmap} as an embodied
characteristic of the geometry of space-time. A similar
investigation has been applied to the case of
electromagnetism\cite{Prezas,beil} together with some recent progress in
gravity, cosmology and fluid dynamics\cite{StavJPh,phoevos,paninf}. We consider the
geodesics of the 4-dimensional space-time  to be produced by a
Lagrangian identified to be the Randers-type metric function(the
greek indices belong to  $\{0,1,2,3\}$ and the latin ones to
$\{1,2,3 \}$)
\begin{eqnarray}
F(x,y)= &\sigma(x,y)+\phi(x)\hat{k}_{\alpha}y^{\alpha} \label{FR}\\
\sigma(x,y)= &\sqrt{a_{\kappa\lambda}(x)y^{\kappa}y^{\lambda}} \label{sigma}
\end{eqnarray}
where $a_{\kappa\lambda}(x)$ is the Robertson-Walker metric defined as
\begin{eqnarray}
a_{\kappa\lambda}(x)=diag(1,-\frac{a^2}{1-kr^2},-a^2r^2,-a^2r^2\sin^2\theta).
\end{eqnarray}
where $k=0,\pm 1$ for a {\it flat, closed} and {\it hyperbolic}
geometry respectively. The spatial coordinates are comoving and
the time coordinate represents the proper time measured by the
comoving observer. The vector $y^{\mu}=\frac{dx^{\mu}}{ds}$
represents  the tangent 4-velocity of a comoving observer along a
preferred family of worldlines(fluid flow lines) in a locally anisotropic
universe;the arclength parameter $s$ stands for the proper time.We
proceed by considering the natural Lorentzian units i.e. $c=1$. If we
fix the direction $y=\dot{x}$ then $\sigma(x,\dot{x})=1$.The
vector field
\begin{eqnarray}
u_{\alpha}(x)=\hat{k}_{\alpha}\phi(x)
\end{eqnarray}
stands for a weak primordial vector field $|u_{\alpha}|\ll 1$
incorporated to the geometry of space-time as an intrinsic
characteristic. This field would most naturally be expected to
point in the same direction with the tangent vectors of the fluid
flow lines\cite{Grav}. As a result it will have only a time like
component which can be expressed as a function of the proper time
$u_a=\left(u_0,0,0,0\right)$. Important information about the
anisotropy is encoded into the component $u_0(t)$\cite{Grav}  . We consider a
linearized variation of anisotropy, therefore the approximation
\begin{eqnarray}
\phi(x)\approx \phi(0)+\partial_{\mu}\phi(0)x^{\mu}\label{linphi}
\end{eqnarray}
is valid for small $x$.

\section{ The choice of the connection $A^{\kappa}_{\lambda\mu}(x)$ and the curvature.}

The metric of the Finsler space can be directly calculated from
the metric function $F$. Since
$f_{\mu\nu}(x,y)=\frac{1}{2}\frac{\partial^2 F^2}{\partial
y^{\mu}\partial y^{\nu}}(x,y)$ we derive
\begin{eqnarray}
f_{\mu\nu}=g_{\mu\nu}+\frac{1}{4\sigma}(u_{\mu}y_{\nu}+u_{\nu}y_{\mu})-\frac{\beta}{\sigma^3}y_{\mu}y_{\nu}+u_{\mu}u_{\nu}\label{fins metric}
\end{eqnarray}
where
\begin{eqnarray}
g_{\mu\nu}(x,y)=\frac{F}{\sigma}(x,y)a_{\mu\nu}(x)\label{fnsg}
\end{eqnarray}
and
\begin{eqnarray}
\beta(x,y)=\phi(x)\hat{k}_{\alpha}y^{\alpha}=u_{\alpha}(x)y^{\alpha}\label{beta}
\end{eqnarray}

Under the weak field assumption  we can approximate the Finslerian
metric (\ref{fins metric})as a perturbation of the FRW metric since in General
Relativity a weak vector field in a space (e.g. primordial
magnetic field) can be treated as first order perturbation of the
Riemann metric tensor. The metric is considered to have signature
$(+,-,-,-)$ for any $(x,y)$. The square of the length of an
arbitrary contravariant vector $X^{\mu}$ is to be defined
$|X|^2=f_{\mu\nu}(x,y)X^{\mu}X^{\nu}$. The connection components
of the metric are given by (\ref{finsg}).

In many cases we consider a convenient Finsler metric to
approximate the gravitational theories\cite{RundFB,Asanov}. This metric is
connected to a Riemannian one, $r_{\mu\nu}(x)$ referred as {\it
osculating Riemannian metric}\cite{RundFB}
\begin{eqnarray}
r_{\mu\nu}(x)=f_{\mu\nu}(x,y(x))
\end{eqnarray}
with the following Christoffel components
\begin{eqnarray}
r^{\kappa}_{\lambda\mu}(x)=\gamma^{\kappa}_{\lambda\mu}(x,y(x)) +C^{\kappa}_{\mu\rho}(x,y(x))\frac{\partial y^{\rho}}{\partial x^{\lambda}}(x)+ \nonumber \\
 +C^{\kappa}_{\lambda\rho}(x,y(x))\frac{\partial y^{\rho}}{\partial x^{\mu}}(x)-g^{\kappa\sigma}(x,y(x))C_{\lambda\mu\rho}(x,y(x))\frac{\partial y^{\rho}}{\partial x^{\sigma}}(x) \label{osccon}
\end{eqnarray}
thus the equation of geodesics is given by
\begin{eqnarray}
\frac{d^2x^\mu}{ds^2}+r^{\mu}_{\rho\sigma}(x)y^{\rho}y^{\sigma}=0.
\label{geodosc}
\end{eqnarray}
Under the assumption that the vector field $y^{\alpha}$ satisfies
the relation $y^{\mu}_{;\nu}=0$  the Finslerian $\delta-$covariant
derivative and the Cartan's covariant derivative of an arbitrary
vector field $X^{\alpha}(x)$ are equal\cite{RundFB,Asanov}(see Appendix B)
\begin{eqnarray}
X^{a}_{;\beta}(x,y(x))=X^{a}_{|\beta}(x,y(x)).
\end{eqnarray}
The Cartan's torsion tensor can be easily deduced from (\ref{FR})
and (\ref{cartan}) the full expression is\cite{phoevos}

\begin{eqnarray}
C_{\mu\nu\lambda}= \frac{1}{2}\left\{ \frac{1}{\sigma}\mathcal{S}_{ (\mu\nu\lambda) }(a_{\mu\nu}u_{\lambda})
-\frac{1}{\sigma^3}\mathcal{S}_{ (\mu\nu\lambda) }(y_{\mu}y_{\nu}u_{\lambda})
-\frac{\beta}{\sigma^3}S_{ (\mu\nu\lambda) }(a_{\mu\nu}y_{\lambda})  \right\} \label{Cartan_fe}
\end{eqnarray}
where $\mathcal{S}_{\mu\nu\lambda}$ denotes the sum over the
cyclic permutation of the indices. Every single term of
(\ref{Cartan_fe}) is proportional to the components of the field
$u_{\alpha}$ thus $C_{\mu\nu\lambda}\approx 0$ under the condition
$|u_{\alpha}|\ll 1$ and then we can drop all the torsion dependent
terms in (\ref{osccon}). Therefore the  approximation for the
Christoffel components becomes
\begin{eqnarray}
A^{\kappa}_{\lambda\nu}(x) \approx \gamma^{\kappa}_{\lambda\nu}(x,y(x))\label{As}
\end{eqnarray}
where $A^{\kappa}_{\lambda\nu}(x)$ represent the osculating affine
connection coefficients. The affine curvature  tensor associated
with the proper choice of the connection coefficients  $A
^{\kappa}_{ \lambda\nu}$ gives directly the curvature  which is
associated with the commutation relations of the
$\delta$-derivatives
\begin{eqnarray}
L^{\kappa}_{\lambda\mu\nu}= A ^{\kappa}_{
\lambda\nu,\mu}-A^{\kappa}_{\lambda\mu,\nu}+A
^{\rho}_{\lambda\nu}A
^{\kappa}_{\rho\mu}-A^{\rho}_{\lambda\mu}A^{\kappa}_{\rho\nu}.
\end{eqnarray}
The Ricci tensor is given by
\begin{eqnarray}
L_{\mu\nu}=  L^{\alpha}_{\mu\alpha\nu}
\end{eqnarray}
and the scalar curvature
\begin{eqnarray}
L=  f^{\mu\nu}L_{\mu\nu}.
\end{eqnarray}
The inverted metric $f^{\mu\nu}$ is calculated in\cite{phoevos}. The
components of the Ricci tensor can be simplified due to
 the conditions
\begin{eqnarray}
\ddot{u}_0\approx 0\label{zmn} \\
{\dot{u}}_0^2\approx 0 \label{zsqr}.
\end{eqnarray}
The condition (\ref{zmn}) is valid since $\phi(x)$ can be written
at the linear form (\ref{linphi}) together with (\ref{zsqr}) where
we have considered $\dot{u}$ very small at the first stages of a
highly accelerated expanding universe\cite{Peebles}. We arrive then at
the following nonzero components
\begin{equation}
\begin{array}{rl}
L_{00}=& 3\left(\ddot{a}/a+3/4\dot{a}/a\dot{u}_0\right) \\
L_{11}=&  -\left( a\ddot{a}+2\dot{a}^2+2k+11/4a\dot{a}\dot{u}_0 \right)/(1-kr^2)  \\
L_{22}=& -\left(a\ddot{a}+2\dot{a}^2+2k+11/4a\dot{a}\dot{u}_0 \right)r^2 \\

L_{33}=& -\left( a\ddot{a}+2\dot{a}^2+2k+11/4a\dot{a}\dot{u}_0  \right)r^2\sin^2\theta
\end{array}\label{Ls}
\end{equation}
The geodesic deviation equation in the case of a perfect fluid
along the neighboring world lines can be generalized within the
Finslerian framework ($\xi^{\mu}$ is  the deviation
vector)\cite{Stavdg,MSRI}
\begin{eqnarray}
\frac{\delta^2 \xi^{\mu}}{\delta s^2}+L^{\mu}_{\nu\rho\sigma}y^{\nu}y^{\rho}\xi^{\sigma}=0\label{dg}
\end{eqnarray}
where the operator $\frac{\delta}{\delta s}$ denotes the Finslerian $\delta-$connection along the geodesics.

Within a Finslerian space-time framework the concept of constant
curvature $K$ is formulated by\cite{phoevos,RundFB}
\begin{eqnarray}
L_{\kappa\lambda\mu\nu}=K(f_{\kappa\mu}f_{\lambda\nu}-f_{\kappa\nu}f_{\lambda\mu}) \label{constcurv}
\end{eqnarray}

\section{Einstein's field equations for an anisotropic universe}

\subsection{The energy-momentum tensor and the Friedman-like equation}

The energy-Momentum tensor of a Finslerian perfect fluid for a
comoving observer\cite{StavJPh,Asanov,Peack} is defined to be
\begin{eqnarray}
T_{\mu\nu}(x,y(x))=(\mu+P)y_{\mu}(x)y_{\nu}(x)-Pf_{\mu\nu}(x,y(x))\label{Tmn}
\end{eqnarray}
where $P\equiv P(x), \mu\equiv \mu(x)$ is the pressure and the
energy density of the cosmic fluid respectively. The vector
$y^{\alpha}=\frac{dx^{\alpha}}{d\tau}$ is the 4-velocity of the
fluid since $y^{\alpha}=(1,0,0,0)$ with respect to comoving
coordinates.Thus $T_{\mu\nu}$ becomes($T_{\mu\nu}=diag( \mu,
-Pf_{ij} )$ in matrix form)\cite{tsagas2,Ohan,Carol,Grav}
\begin{equation}
\begin{array}{rl}
T_{00}=& \mu  \\
T_{ij}=& -Pf_{ij} \\
     T=& T^{\mu}_{\mu} \\
    \:=& f^{00}\mu -3P
\end{array}\label{Tmn_comp}
\end{equation}
The substitution of (\ref{Tmn}) to the field equations
\begin{eqnarray}
L_{\mu\nu}=8\pi G\left( T_{\mu\nu}-\frac{1}{2}Tg_{\mu\nu} \right)
\end{eqnarray}
implies the following equations at the weak field limit

\begin{eqnarray}
\frac{\ddot{a} }{a} +\frac{3}{4}\frac{\dot{a} }{a}\dot{u}_0 = -&\frac{4\pi G}{3}(\mu +3P) \label{EIN1}\\
\frac{\ddot{a} }{a}
+2\frac{\dot{a}^2}{a^2}+2\frac{k}{a^2} +\frac{11}{4}\frac{\dot{a} }{a}\dot{u}_0 =&
4\pi G(\mu -P)\label{EIN2}
\end{eqnarray}
\newline
after subtracting (\ref{EIN1}) from (\ref{EIN2}) we obtain the Friedman-like equation
\begin{eqnarray}
\left(\frac{\dot{a}}{a}\right)^2+ \frac{\dot{a}}{a}z_t= \frac{8\pi
G}{3}\mu-\frac{k}{a^2} \label{AFRD}
\end{eqnarray}
where we define $z_t$ as
\begin{eqnarray}
z_t=\dot{u}_0 .
\end{eqnarray}
The quantity $z_t$ is a constant since we consider a linear
approach for the $\phi(x)$ by (\ref{linphi}).  The previous
equation is similar to the one derived from the Robertson-Walker
metric in the Riemannian framework, apart from the extra term
$\frac{\dot{a}}{a}z_t$.We associate this extra term to the present
Universe's anisotropy.In case we study Finslerian models with a
cosmological constant\cite{diInv} the field equations
(\ref{EIN1}),(\ref{EIN2}) can be given in the following form
\begin{eqnarray}
\frac{\ddot{a} }{a} +\frac{3}{4}\frac{\dot{a} }{a}\dot{u}_0  = -&\frac{4\pi G}{3}(\mu +3P)+\frac{\Lambda}{3} \label{LEIN1}\\
\frac{\ddot{a} }{a}
+2\frac{\dot{a}^2}{a^2}+2\frac{k}{a^2}+\frac{11}{4}\frac{\dot{a} }{a}\dot{u}_0  =&
4\pi G(\mu -P)+\Lambda\label{LEIN2}
\end{eqnarray}
and we end up with the equation of motion
\begin{eqnarray}
\left(\frac{\dot{a}}{a}\right)^2+ \frac{\dot{a}}{a}z_t= \frac{8\pi
G}{3}\mu-\frac{k}{a^2}+\frac{\Lambda}{3}. \label{LAFRD}
\end{eqnarray}

\subsection{The  parameter $z_t$ and the weak linearized anisotropy}
The physical quantity $z_t$ describes the variation of anisotropy
which evolves linearly due to (\ref{linphi});it depends upon the
scalar $\phi(x)$ which is the only quantity of the Lagrangian that
gives us insight about the evolution of anisotropy.The parameter
$z_t$ is measured by the Hubble's units as (\ref{AFRD}) implies.
It is significant that $z_t$ depends on the geometrical properties
of the Finslerian space-time manifold.Indeed, the component
$C_{000}$ can be directly calculated from (\ref{Cartan_fe}) as

\begin{eqnarray}
C_{000}=\frac{u_0}{2}
\end{eqnarray}
and after differentiating with respect to proper time we lead to {\it the direct dependence of
$z_t$ on the Cartan torsion component $C_{000}$ }
\begin{equation}
z_t=2C_{000,0}\label{zet c}
\end{equation}
hence the variation of anisotropy  is closely related to the
variation of the Cartan torsion tensor as an intrinsic object of
the Finslerian space-time.

\section{The cosmological anisotropic parameters}
We list the main anisotropic parameters constructed within the
Finslerian framework\cite{StavJPh,Stavcong,phoevos,Asanov}

\subsection{The anisotropic scale factor $\tilde{a}(\upsilon(s))$ and the Hubble parameter $\tilde{H}$}
The anisotropic scale factor is defined along each world line.
$S(s)$ is the length scale introduced in \cite{MV,StavJPh} defined as
\begin{eqnarray}
S(s)=\tilde{a}(\upsilon(s))
\end{eqnarray}
where $\upsilon(s)$ is the tangent
vector field along the world lines. The anisotropic Hubble
parameter $\tilde{H}$ is given by the
relation($\dot{S}=\frac{\partial\tilde{a}}{\partial
\upsilon^{\mu}}\dot{\upsilon}^{\mu}(s)$)
\begin{eqnarray}
\tilde{H}=\frac{\dot{S}}{S}=\frac{1}{3}\tilde{\Theta}
\end{eqnarray}
The term $\tilde{\Theta}$ is the expansion in the Finslerian space-time expressed as
\begin{eqnarray}
\tilde{\Theta}=y^{\mu}_{;\mu}-C^{\lambda}_{\mu\lambda}\dot{y}^{\mu}\label{theta}
\end{eqnarray}
The anisotropic Hubble parameter can be computed from (\ref{AFRD})
\begin{eqnarray}
\tilde{H}^2 & =\frac{8\pi G}{3}\mu-\frac{k}{a^2}\nonumber \\
            & =H^2+Hz_t \label{AHB}
\end{eqnarray}
thus  Hubble's units have to be attributed to $z_t$.Since $\tilde{H}^2>0$ we should fix $H^2\gg |Hz_t|$
Therefore the lower limit of $z_t$ is $-H$,
\begin{equation}
z_t\geq -H .
\end{equation}
This result is in agreement with
the fixing of a similar parameter in\cite{Mart_NEB,Mart_MG,Mart_lcdm} for self-accelerated  brane-world
cosmology.Since the parameter $z_t$ is related to
the variation of anisotropy we expect it to have negative
sign({\it self accelerating universe}) as it may control a transition of the
universe from a state of anisotropy to a smoother isotropic phase\cite{bogoslovski}.

\subsection{The density and the deceleration parameter}
The density parameter can be defined with respect to the parameter $\tilde{H}$
\begin{eqnarray}
\tilde{\Omega}_{\mu}=\frac{8\pi G}{3 \tilde{H}^2}\mu=\frac{\mu}{\tilde{\mu}_{crit}} \label{omega}
\end{eqnarray}
where
\begin{eqnarray}
\tilde{\mu}_{crit}=\frac{3\tilde{H}^2}{8\pi G}.\label{mcr}
\end{eqnarray}
The deceleration parameter is defined in terms of the anisotropic scale factor $S(s)$
\begin{eqnarray}
\tilde{q}=-S\ddot{S}/\dot{S}^2.\label{tq}
\end{eqnarray}
The Friedman-like equation can also be rewritten in the form
\begin{eqnarray}
\tilde{\Omega}_{\mu}-1=k/(\tilde{H}^2a^2)
\end{eqnarray}
If $\mu<\tilde{\mu}_{crit}$ then $\tilde{\Omega}_{\mu}<1$ or $k<0$
{\it (open universe)}.If $\mu>\tilde{\mu}_{crit}$ then
$\tilde{\Omega}_{\mu}>1$ or $k>0$ {\it (closed universe)}.The latter
case $\mu=\tilde{\mu}_{crit}$ corresponds to
$\tilde{\Omega}_{\mu}=1$ or $k=0$ { \it (flat universe) }. Thus
the values of $z_t$ influence the type of spatial curvature. In
case we need to express $\tilde{H}$ in terms of the redshift, the
present value of the Hubble parameter $H_0$ and the
$\Omega_{M_0}$,$\Omega_{\Lambda_0}$ we insert\cite{Peack,Peebles,Carol}
\begin{equation}
H(z)= H_0E(z) \label{HEz}
\end{equation}
into (\ref{AHB}) where the quantity $E(z)=H(z)/H_0$ is given by

\begin{equation}
E(z)= \sqrt{ \Omega_{ z_t } }+\sqrt{ \Omega_{ z_t }+\Omega_{M_0}(1+z)^3+\Omega_{K_0}(1+z)^2+\Omega_{\Lambda_0} } \label{FRDz}
\end{equation}
and the parameter $\Omega_{z_t}$ is defined by
\begin{equation}
\Omega_{ z_t }=\frac{z_t^2}{4H_0^2}. \label{omegazt}
\end{equation}
Therefore we are dealing with an expression for $\tilde{H}(z)$
which depend up on the anisotropic parameter $z_t$, the redshift $z$ and
the $\Omega$'s. \newline\newline
{\it {\large The Friedman equations in terms of $\Omega$'s} }\newline
The Friedman equations can be expressed in turns of the density parameters.The $\Omega$'s
give back some useful insight about the unknown parameter $z_t$ and enable us to test
if our cosmological model fits to the current data(e.g.WMAP data).Indeed, the equations of
motion (\ref{AFRD}),(\ref{LAFRD}) are reduced to the form
\begin{equation}
1= \Omega_{M_0}+\Omega_{K_0}+\Omega_{\Lambda_0}-z_t/H_0. \label{ztest}
\end{equation}
In case we are interested in manipulating the Friedman equation for a specific value of the
redshift $z$ we can make use of (\ref{FRDz}).It is difficult to distinguish the last term of (\ref{ztest}) from GR
 dark energy, since both terms accelerate the universe. If the self-accelerating term $z_tH$ dictates over
the expansion, the cosmological constant in (\ref{ztest}),(\ref{FRDz}) has to
vanish \cite{Mart_MG}.

\subsection{The continuity equation}
The energy density can be calculated by direct integration of the continuity equation $T^{\mu}_{0;\mu}=0$. A proper
manipulation of (\ref{Tmn}) and (\ref{As}) leads to the equation
\begin{eqnarray}
T^0_{0,0}+\left(\dot{a}/a+ \dot{u}_0/2 \right)\left(3T^0_0-T^i_i\right)=0
\end{eqnarray}
which can be simplified to the form($c=1$)
\begin{eqnarray}
\dot{\mu}+3\dot{a}/a\left(\mu+P\right) +\dot{u}_0/2\left(3P+2\mu\right)=0 \label{cont}
\end{eqnarray}
since we have applied the approximations
\begin{equation}
\begin{array}{rl}
f^{00}=& 2/(2+u_0+2u_0^2)\approx 1 \\
f^{00}_{,0}=& -2\dot{u}_0(1+4u_0)/(2+u_0+2u_0^2)^2  \\
T^0_{0,0}=& f^{00}\dot{\mu}+f^{00}_{,0}\mu \approx \dot{\mu}-\mu\dot{u_0}/2 \\
3T^0_0-T^i_i \approx&  3\mu+3P.
\end{array}
\end{equation}
A perfect fluid relevant to cosmology obeys the equation of state
\begin{eqnarray}
P=w\mu  \label{eqst}
\end{eqnarray}
where $w=0$ for a matter dominated universe and $w=1/3$ for a radiation dominated universe.The substitution of (\ref{eqst})
to (\ref{cont}) leads to
\begin{eqnarray}
\dot{\mu}/\mu=-3(1+w) \dot{a}/a+ \dot{u}_0/2(2+3w) \label{ODEcon}
\end{eqnarray}
and the integration of the differential equation (\ref{ODEcon}) implies
\begin{eqnarray}
\mu\propto a^{-3(1+w)}\exp\left(-u_0\left( 3w+2 \right)/2\right)
\end{eqnarray}
therefore
\begin{equation}
\mu\propto\left\{
\begin{array}{ll}
a^{-3}\exp(-u_0) & \mbox{{\it matter dominated universe}}\\
a^{-4}\exp(-3u_0/2)      & \mbox{{\it radiation dominated universe}}
\end{array}\right.\label{mu}
\end{equation}
This asymptotic behavior indicates that the weak anisotropy {\it affects homogeneity}.

\subsection{The look back time at the presence of weak anisotropy}
Given the definition of the anisotropic Hubble parameter
(\ref{AHB}) we can generalize the concept of the lookback time
$t_0-t_*$ in\cite{Carol}
\begin{eqnarray}
\tilde{t}_0-\tilde{t}_*=\int_0^{z_*}\frac{ dz^{\prime} }{ (1+z^{\prime})\tilde{H}(z^{\prime}) }
\end{eqnarray}
where $t_0$ is the age of the universe today and $t_*$ is the age
of the universe when the redshift was $z=z_*$.Therefore  the
general expression for the world time evolution should be\cite{Peebles}
\begin{eqnarray}
\tilde{t}(z)=\int_{z}^{\infty}\frac{ dz^{\prime} }{ (1+z^{\prime})\tilde{H}(z^{\prime}) }
\end{eqnarray}
since $z\rightarrow\infty$ at the start of the universe.Taking into account the weak anisotropy scenario
and the redshift expression of the Hubble parameter (\ref{HEz}) into account we deduce the small $z_t$
expansion for the worldtime today $\tilde{t}_0$($z=0$)
\begin{eqnarray}
\tilde{t}_0=t_0+\sum_{k=1}^{\infty}T_{k}z_t^k
\end{eqnarray}
where
\begin{eqnarray}
T_k=H_0^{-k-1}\left(
\begin{array}{c}
-1/2 \\
 k
\end{array}
\right)\int_0^{\infty}\frac{dz}{ (1+z)E(z)^{k+1} }.
\end{eqnarray}

\subsection{The Finslerian Raychaudhuri equation}
The generalization of Raychaudhuri's equation has been given by
the following formula\cite{StavJPh}
\begin{eqnarray}
\dot{\tilde{\Theta}}=- \frac{1}{3}\tilde{\Theta}^2-
\tilde{\sigma}_{\mu\nu}\tilde{\sigma}^{\mu\nu}
+\tilde{\omega}_{\mu\nu}\tilde{\omega}^{\mu\nu}-4\pi
G(\mu+3P)+(y^{\mu}_{;\kappa}y^{\kappa})_{;\mu}\label{raych}
\end{eqnarray}
where
$\tilde{\sigma}^2=\tilde{\sigma}_{\mu\nu}\tilde{\sigma}^{\mu\nu}$,
$\tilde{\omega}^2=\tilde{\omega}_{\mu\nu}\tilde{\omega}^{\mu\nu}$
are the Finslerian {\it shear} and {\it vorticity} respectively,
which are defined in\cite{RundFB,Asanov}and refer to a perfect fluid. The
Eq.(\ref{raych}) is a direct application  of the Finslerian Lie
derivative for dust-like matter developed in\cite{StavLie}. It includes the anisotropic gravitational influence of
the matter along the worldlines which is expressed by the tidal
force of the field
\begin{eqnarray}
L_{\mu\nu}y^{\mu}y^{\nu}=4\pi G(\mu +3P).
\end{eqnarray}
Using the expressions  (\ref{tq}),(\ref{theta}) we produce the linearized
Raychaudhuri equation( {\small $\tilde{q}=
-\frac{\ddot{\tilde{a}}}{ \tilde{a} }\tilde{H}^{-2}$ } where
$\tilde{H}$ depends on $a,\dot{a}$ in virtue of (\ref{AHB})  )
\begin{eqnarray}
\frac{1}{3}\tilde{\Theta}^2\tilde{q}=4\pi G(\mu +3P)\frac{H}{\tilde{H}}+
f(a,\dot{a},\ddot{a},z_t)\label{LRay}
\end{eqnarray}
where the function $f(a,\dot{a},\ddot{a},z_t)$ is defined as
\begin{equation}
\begin{array}{rl}
f(a,\dot{a},\ddot{a},z_t)=& -3\tilde{H}^2-\frac{3}{2}\frac{H}{\tilde{H}}(\tilde{H}^2+H^2)+ \nonumber \\
                         +& \frac{1}{\tilde{H}}\left[2\pi G(\mu+3P)+3z_tH\right]z_t.
\end{array}\label{fraych}
\end{equation}
Thus we can  expand for small values of $z_t$ and conclude that
\begin{eqnarray}
\frac{1}{3}\tilde{\Theta}^2\tilde{q}=4\pi G(\mu+3P)+3\frac{\dot{a}}{a}z_t-\frac{3}{8}\frac{\ddot{a}(t)a(t)-\dot{a}(t)^2}{\dot{a}(t)^2}z_t^2+O(z_t^3)
\end{eqnarray}
The sign of the right hand side of (\ref{LRay}) determines the state of
expansion.If the inequality $f(a,\dot{a},\ddot{a},z_t)<0$ is valid
then the term $f(a,\dot{a},\ddot{a},z_t)$ contributes to the
acceleration of the universe(the field assist inflation), whereas if
$f(a,\dot{a},\ddot{a},z_t)>0$ it will slow the expansion down(the
inflation domination must be longer to accelerate the
universe).This specific effect is due to the kinematical reaction of the
geometry of the spatial hypersurfaces, rather than an attempt to
suppress inflation(it can be considered as an essential ingredient
of the {\it Finslerian ansatz})\cite{tsagas2,RundFB}.

\section{ Einstein field equations with anisotropic term}
We assume $z_t$ to be a constant and study the differential
equations (\ref{EIN1}),(\ref{EIN2}) and (\ref{AFRD}) both for the
cases of  matter and radiation dominated universe. We notice that
for $z_t=0$ the field equations reduce to the usual ones
coming from a Riemannian Robertson Walker metric\cite{Ohan,Grav,Wald}. The
whole calculation is done for a homogeneous universe of constant
density $ \mu $.

\subsection{Solution for a matter dominated universe}
The solution for the scale factor in the case of a matter
dominated universe is  derived by the integration of the
Friedman-like equation (\ref{AFRD}) with initial condition
\begin{eqnarray}
a(0)=0
\end{eqnarray}
where $z_t=const$. We set up $t=0$ as the beginning of time
without considering any quantum effects;there are different ways
of handling the initial condition for example setting
$a(t_{Pl})=0$ or considering the scale factor after the Plank
scale(e.g. see \cite{Maartens}.The Eq. (\ref{AFRD}) is simplified if we
insert the parameter (conformal proper time)\cite{Ohan,Grav}
\begin{eqnarray}
\eta=\int_0^t\frac{d\omega}{a(\omega)}
\end{eqnarray}
which measures the arc in rad traveled along by a photon on a
sphere of radius  $a(t)$.We study the asymptotics of the scale factor
\begin{eqnarray}
a(t)\equiv a(t(\eta))=\bar{a}(\eta)\rightarrow 0
\end{eqnarray}
where $cdt=ad\eta$. We consider a homogeneous universe of constant
density $\mu$ calculated as \cite{Grav}
\begin{eqnarray}
\mu=\frac{M}{V}=\frac{M}{2\pi^2a^3}.
\end{eqnarray}
We assume that the universe would take over the same volume as in the closed case ($k=1$)
since (\ref{mu}) is valid for the asymptotics of $\mu$. The parameter $k$ determines the kind of spatial geometry $k=0,-1,+1$ ({\it flat, open, closed} universe
respectively).Since $\frac{\dot{a}}{a}=\bar{a}^{-2}\dot{\bar{a}} $ Eq.(\ref{AFRD}) becomes
\begin{eqnarray}
\left(\dot{\bar{a}}+\frac{z_t}{2}\bar{a}^2\right)^2=  \frac{z_t^2}{4}\bar{a}^4-k\bar{a}^2+\frac{4GM}{3\pi}\bar{a} \label{FRDint}
\end{eqnarray}
We study a Universe that accelerates very fast at its early stages thus we can accept
 $\dot{\bar{a}}+\frac{z_t}{2}\bar{a}^2>0$. The velocity of expansion $\dot{a}$ takes on very
large values. The Eq.(\ref{FRDint}) can be integrated directly for
all the values of $k$. It is more convenient to substitute\cite{Ohan}
\begin{eqnarray}
a_*=\frac{2GM}{3\pi }
\end{eqnarray}
and  the separable Eq.(\ref{FRDint}) leads to
\begin{eqnarray}
t=-\frac{z_t}{2}\left\{
\int_0^a\frac{\sqrt{x^4-4k/z_t^2x^2+8a_*/z_t^2x}}{kx-2a_*}dx
+\int_0^a\frac{x^2}{kx-2a_*}dx\right\}\label{integr}
\end{eqnarray}
together with the initial condition
\begin{eqnarray}
a(0)= 0.
\end{eqnarray}
\newline
{\bf I) calculation for $k=0$}
\newline
We fix $k=0$ at (\ref{integr}) and find
\begin{equation}
\begin{array}{rl}
t=& \frac{z_t}{2}\left[ \frac{a^3}{6a_*}+I_0 \right]  \\
I_0=& \frac{1}{2a_*}\int_0^a \sqrt{ x^4+8a_*/z_t^2x }dx
\end{array}
\end{equation}
we expand for a small $z_t$ and arrive at the solution
\begin{eqnarray}
t=\frac{\sqrt{2}}{3\sqrt{a_*}}a^{3/2}+1/(12a_*)a^3z_t/c+O(z_t^2)
\end{eqnarray}
\newline
{\bf II) calculation for $k=-1$}\newline
A direct integration of(\ref{integr}) leads to
\begin{eqnarray}
t=-\frac{z_t}{4}\left[a^2/2+2a_*a+4a_*^2\log|a-2a_*| +I_1 \right]
\end{eqnarray}
where
\begin{eqnarray}
I_1=\int_0^a \frac{ \sqrt{x^4-4/z_t^2 x^2+8a_*/z_t^2x} }{x-2a_*}dx
\end{eqnarray}

after expanding for small $z_t$
\begin{equation}
\begin{array}{rl}
t=& \sqrt{a(a+2a_*)}-a_*\log\left(1+a/a_*+\sqrt{a(a+2a_*)}/a_*  \right)  \\
  -& \left(a^2/4+a_*a+2a_*^2\log|a-2a_*|  \right)z_t+O(z_t^2).
\end{array}
\end{equation}
\newline
{\bf III) calculation for $k=+1$}
\newline
The calculation is the same as the previous case
\begin{eqnarray}
t=\frac{z_t}{2}\left[a^2/2-2a_*a+4a_*^2\log|a+2a_*| +I_{-1} \right]
\end{eqnarray}
where
\begin{eqnarray}
I_{-1}=\int_0^a \frac{ \sqrt{x^4+4/z_t^2 x^2+8a_*/z_t^2x}
}{x+2a_*}dx
\end{eqnarray}
and after expanding for small $z_t$ we obtain the solution
\begin{equation}
\begin{array}{rl}
t=&-\sqrt{a(2a_*-a)}+a_*\arccos\left(1-a/a_*\right)+ \\
  +&\left(a^2/4-a_*a+2a_*^2\log|a+2a_*|\right)z_t+O(z_t^2).
\end{array}
\end{equation}

The leading term of the solutions for all k represents the solution given by the field's equations
of the Robertson-Walker metric\cite{Ohan}.A small $a$ expansion gives the asymptotic behavior $t\sim a^{3/2}$ or equivalently
\begin{eqnarray}
a\sim t^{2/3}.
\end{eqnarray}

\subsection{Solution for a radiation dominated universe}
The solution for a radiation dominated universe can be deduced by  inserting the equation of state
 $P=\frac{1}{3}\mu_{rad}$ into (\ref{EIN1}) and after adding this to (\ref{EIN2}) we end up
with the equation
\begin{eqnarray}
\frac{d}{dt}(a\dot{a})=-k-\frac{7}{4}a\dot{a}z_t \label{2ndODE}
\end{eqnarray}
which we integrate and find
\begin{eqnarray}
\dot{a}a+\frac{7}{4}z_ta^2=-kt+C_{1} \label{auxilliary}
\end{eqnarray}
If we substitute $z_t=0$ to (\ref{auxilliary}) we get back the usual solution for a radiation dominated universe $a\propto \sqrt{t}$ for all
values of $k$.If we take into account the initial condition $a(0)=0$  we arrive at the solution
\begin{eqnarray}
a(t)=\frac{4\sqrt{2}}{7z_t}\left\{(C_0z_t+k)\left(1-\exp(-\frac{7}{4}z_t t)\right)-\frac{7}{4}kz_t t \right\}^{1/2}\label{RSL}.
\end{eqnarray}
The expansion of the solution for small $z_t$ is
\begin{eqnarray}
a(t)=\sqrt{t}\left\{ \sqrt{2C_0-kt}+\frac{7t(kt-3C_0)}{6 \sqrt{4C_0-2kt} }\cdot z_t+O(z_t^2) \right\} \label{arexp}
\end{eqnarray}

\subsection{Solution for the de-Sitter model}
The de-Sitter model for an empty anisotropic universe constructed
in\cite{StavJPh} leads to the equation of motion($\mu=0$ in (\ref{LAFRD}))
\begin{eqnarray}
\left(\frac{\dot{a}}{a}\right)^2+ \frac{\dot{a}}{a}z_t=-\frac{k}{a^2}+\frac{\Lambda}{3} \label{EMdS}
\end{eqnarray}
Since $\tilde{H}^2$ in (\ref{EMdS}) can be written as $\tilde{H}^2=\left(\frac{\dot{a}}{a}\right)^2+ \frac{\dot{a}}{a}z_t\geq 0$,
the cosmological constant $\Lambda$ is restricted by the inequality $\Lambda\geq 3K$ where $K=k/a^2$ is the
curvature of the space.\newline
{\bf I) calculation for  $k=0$}\newline
The case of zero curvature can be integrated to give
\begin{eqnarray}
a(t)=const\times\exp\left(-z_t/2t\right)\exp\left[\left(\Lambda/3+z_t^2/4\right)^{1/2}t\right]\label{DSzt}
\end{eqnarray}
The solution converges to the one without anisotropy if we let
$z_t\rightarrow 0$\cite{diInv}. \newline {\bf II) calculation for the
special case $\Lambda=3K$}\newline The field equations for a space
of constant spatial curvature(maximal symmetry) and an empty
universe($T_{\mu\nu}=0$) imply the condition\cite{StavJPh,Carol}
\begin{eqnarray}
\Lambda=3K
\end{eqnarray}
hence $\Lambda$ can be substituted to the equation of motion (\ref{EMdS})

\begin{eqnarray}
\left(\frac{\dot{a}}{a}\right)^2+ \frac{\dot{a}}{a}z_t=0
\end{eqnarray}
and the solution for the scale factor is
\begin{eqnarray}
a(t)=const\times\exp\left(-z_tt\right).
\end{eqnarray}
In both cases the constant of integration can be absorbed into $a(t)$ if we choose the right
scale(e.g.$a(0)=1$).

\subsection{A model for inflation with anisotropy}

The idea of inflation can be incorporated into the model of
Friedman-like equations with weak anisotropy if we introduce the
vacuum energy density of a scalar field $V_0$ to the energy
density $\mu$.Indeed, we consider $G\mu=V_0/m_{Pl}^2$, i.e.
$\mu=V_0$ thus (\ref{AFRD}) implies\cite{Carol,Liddle,diInv}
\begin{eqnarray}
\dot{a}^2+a\dot{a}z_t= \frac{8\pi}{3}\frac{V_0}{m_{Pl}^2}a^2-k. \label{FRDinf}
\end{eqnarray}
Since we work at an inflationary phase the size of the scale factor is such that the term
$ \frac{8\pi}{3}\frac{V_0}{m_{Pl}^2}a^2$ dominates over $k$ hence we can neglect $k$ and (\ref{FRDinf})
can be rewritten as
\begin{eqnarray}
\tilde{H}^2=\frac{8\pi}{3}\frac{V_0}{m_{Pl}^2}\label{HubInf}.
\end{eqnarray}
Taking the positive square root of (\ref{HubInf}) we find the scale factor
\begin{eqnarray}
a(t)=const\times\exp\left(-z_t/2t\right)\exp\left[\left(8\pi/3\cdot V_0/m_{Pl}^2+z_t^2/4\right)^{1/2}t\right]
\end{eqnarray}
recovering the de-Sitter solution (\ref{DSzt}) and the expected exponential rate of expansion
for the early inflationary phase of the universe.

\section{Estimation of the  {\it Cosmic Microwave Background Radiation (CMB)}}
The estimation of CMB can be achieved with the aid of
Stefan-Boltzmann's law\cite{Ohan,Mandl}
\begin{eqnarray}
\mu_{rad}c^2=\sigma_{SB}g_*T(z)^4 \label{SB}
\end{eqnarray}
where
\begin{eqnarray}
\sigma_{SB}=\frac{\pi^2k_B^4}{30\hbar^3c^3 }.
\end{eqnarray}
The temperature $T(z)$ is the radiation temperature for a given redshift $z$ and
\begin{eqnarray}
g_*=\sum_{ \mbox{bosons} } g_i+\frac{7}{8}\sum_{ \mbox{fermions} } g_i
\end{eqnarray}
is defined as the sum of the boson and fermion  spin states(e.g. for photons $g=2$, for
neutrinos $g=1$ and for massive particles $g=2s+1$ ).The calculation could be  attainable if
we consider some data for the Hubble parameter $H=\dot{a}/a$ and the anisotropic constant
$z_t$.The radiation dominated solution does not depend on the nature of the spatial geometry
of the universe thus we fix $k=0$ in  (\ref{AFRD}) and obtain $\mu_{rad}$

\begin{equation}
\begin{array}{rl}
\mu_{rad}= & \frac{3}{8\pi G}(H^2+Hz_t)  \\
         = & \frac{3}{8\pi G}\tilde{H}^2
\end{array}\label{MRAD}
\end{equation}
therefore (\ref{SB}),(\ref{MRAD}) yield\cite{Ohan}
\begin{eqnarray}
T(z)=\left\{\frac{1}{\sigma_{SB}g_*}\left(\frac{3}{8\pi G}\tilde{H}^2c^2 \right) \right\}^{1/4}. \label{Tzt}
\end{eqnarray}
\newline
The calculation can be directly derived if we manipulate (\ref{Tzt}) into the form
\begin{eqnarray}
k_B^2T(z)^2= \frac{3\sqrt{5}}{2\sqrt{g_*}\pi^{3/2}} m_{Pl}c^2\hbar \tilde{H}
\end{eqnarray}
where $m_{Pl}$ stands for the Planck mass
\begin{eqnarray}
m_{Pl}=\sqrt{\frac{\hbar c}{G}}.
\end{eqnarray}
Thus we can calculate the temperature $T(z)$ for a given value of the redshift $z$
with the aid of the formula
\begin{eqnarray}
T(z)=T_H(z)\left[1+z_t/H(z)\right]^{1/4}\label{Tz}
\end{eqnarray}
where $T_H(z)$ is the value of the temperature for a given value of the Hubble parameter without
the assumption of weak anisotropy and $H(z)$ is calculated from (\ref{HEz}).Due to the adiabatic
expansion of the universe we can transform the value $T(z)$ at the redshift of CMB to its
present value $T_0$ using the formula($a_{CMB}=1/(1+z_R),z_R\approx 1090$)
\begin{eqnarray}
T_0=a_{CMB}T(z).
\end{eqnarray}

\section{Discussion}
The study of a FRW-model with a weak vector field incorporated in the metric structure of space-time
provides us the extended Friedman-like Eq.(\ref{AFRD}).The contribution of the
variation of anisotropy is expressed by the additional parameter $z_t$  produced by the
Finslerian character of the geometry of space-time. Especially as it is evident from Eq.(\ref{zet c}) $z_t$ has a direct dependence
up on the Cartan torsion component $C_{000}$. We remark that
our present model correspond to the ones studied in \cite{Mart_MG,Mart_lcdm} for a flat universe
due to the correspondence of $z_t$ to $\pm \frac{1}{r_c}$, where $r_c$ is the extra parameter defined
there.The extra parameter $z_t$ appears to compete against the contribution of the cosmological constant
due to (\ref{ztest}).

We perform the model-independent and insensitive to perturbation $S$-test\cite{Wang_AsJ},
where $S$ is the CMB shift parameter\cite{Mart_MG,Mart_lcdm}.We are pretty confident that our model
reproduces the WMAP data for a flat universe since for $S=1.70\pm 0.03$\cite{tywmap} we end up
with $|\Omega_K|\ll 1$.The same result seems to be valid if we apply the baryon acoustic oscillation
peak test for $A=0.469\pm 0.017$\cite{Fair_Goob,Eisenstein}.The procedures of the tests and the
formulas for $A$ and $S$ can be found in \cite{Mart_MG}.A part of a future  work is the investigation
of cosmological perturbations since the Finslerian approach generates deviations from homogeneity and isotropy.
A more fundamental task is the comparison of our model to the data of CMB anisotropies and the
matter power spectrum, strongly connected to the analysis of the density perturbations\cite{KMaart}.

The initial highly compressed, thermal radiation dominated state of anisotro-py  is considered
to be adiabatically transformed to a cooler matter dominated isotropic phase\cite{Grav}.In such
a case where the anisotropy energy is converted to thermal energy and large amount of entropy\cite{Mat_Mis},
a phase transition in the geometric structure can be regulated by the second term of Eq.(\ref{FR}), since we expect 
$\phi(x)$ monotonically decrease as the universe expands, in order to obtain the standard FRW model. 
This ensures negative values for the parameter $z_t$ and provides us a self accelerating cosmological model.

The whole picture of anisotropy directed by a primordial vector field can be locally
incorporated to the anisotropic metric structure of a Finslerian space-time.The
osculation of the Finsler space leads to the construction of a disformal Riemann structure
and can be interpreted as a model of modified gravity.

\section{Acknowledgments}
We wish to thank R.Maartens for the useful comments on the text and
the University of Athens(Special Accounts for Research Grants)for the support to this work.

\appendix

\section{Appendix }
In the following we present some basic elements of Finsler
geometry\cite{newref1,newref2,MSRI,cliff,RundFB}. In 1854 B.Riemann, before arriving at
Riemannian metric was concerned with the concept of a more
generalized metric
\begin{eqnarray}
ds^2=\mathcal{F}(x^1,x^2,...,x^n,dx^1,...,dx^n)
\end{eqnarray}
where $n$ is the dimension of the space.A Finsler structure is
provided by a n-dimensional $C^{\infty }$ manifold $M^{n}$, a
 $C^{\infty}$ function $F\equiv F(x,y)$ defined on the tangent bundle
$\tilde{TM}=TM/\{0\}$, $F:\tilde{TM}\rightarrow R$ that
satisfies the conditions
\begin{equation}
\begin{array}{rl}
(F1) & \mathcal{F}(x,y)>0 \:\: \forall\:\: y\not=0 \\
(F2) & \mathcal{F}(x,py)=p\mathcal{F}(x,y)\:\: \mbox{for any}\:p>0
\end{array}\label{pd_homog}
\end{equation}
where $y$ denotes the directions or velocities on the considered
manifold with the previous coordinates.The metric tensor(Hessian)
\begin{eqnarray}
f_{ij}(x,y)=\frac{1}{2}\frac{\partial^2 \mathcal{F}^2}{\partial
y^i\partial y^j}(x,y) \label{gij}
\end{eqnarray}
is of $rank[(f_{ij})_{i,j}]=n$ and homogeneous of zero degree with
respect to $y$ due to the Euler's theorem.
 The length $s$ of a curve
$C:x^i(t),a\leq t\leq b$ on the manifold is
\begin{eqnarray}
s=\int_a^b \mathcal{F}(x(t),y(t))dt.
\end{eqnarray}
The integral of the length is independent of the parameter if and
only if the condition (F2) is valid. The condition of homogeneity
enables us to define the line element
\begin{eqnarray}
ds=\mathcal{F}(x,dx)
\end{eqnarray}
and the variation of the arclength $\delta\int ds=0$ implies the
Euler-Lagrange equations $\frac{d}{ds}\left( \frac{\partial
\mathcal{F}}{\partial y}(x,y) \right)-\frac{\partial
\mathcal{F}}{\partial x}(x,y)=0$ which represent the geodesics of
the Finsler space. The equation of geodesics then becomes
analogous to the ones of the Riemann space
\begin{eqnarray}
\frac{d^2x^i}{ds^2}+\gamma^i_{jk}y^jy^k=0 \label{geodG}
\end{eqnarray}
where the Christoffel symbols are defined by the usual formula
\begin{eqnarray}
\gamma^i_{jk}(x,y)=\frac{1}{2}f^{ir}(x,y)\left(f_{rj,k}(x,y)+f_{rk,j}(x,y)-f_{jk,r}(x,y)\right)
\label{finsg}.
\end{eqnarray}

The notion of torsion tensor is crucial within the Finsler
Geometry's framework. {\it A Finsler space is a Riemann space if
and only if $C_{ijk}=0$} where $C_{ijk}$ is the torsion tensor
defined by E.Cartan as
\begin{eqnarray}
C_{ijk}=\frac{1}{2}\frac{\partial f_{ij}}{\partial
y^k}\label{cartan}
\end{eqnarray}
Therefore a Finsler space can be treated as a natural
generalization of a Riemann space.

\section{Appendix }
For a Finslerian vector field $X^{\alpha}(x,y(x))$ the $\delta -$
covariant derivative has the form\cite{RundFB,Asanov}
\begin{eqnarray}
X^{a}_{;\beta}(x,y(x))=X^{\alpha}_{,\beta}(x)+\Gamma^{*\alpha}_{\rho\beta}(x,y(x))X^{\rho}(x)
\label{deltaderiv}
\end{eqnarray}
and the Cartan's covariant derivative is given by
\begin{equation}
\begin{array}{rl}
X^{a}_{|\beta}(x,y(x))= & X^{\alpha}_{,\beta}(x)-\frac{\partial X^{\alpha}}{\partial y^{\rho}}(x,y(x))G^{\rho}_{\beta}(x,y(x))+ \\
                      + & \Gamma^{*\alpha}_{\rho\beta}(x,y(x))X^{\rho}(x).
\end{array} \label{cdcov}
\end{equation}
The $\Gamma^{*\kappa}_{\lambda\mu}$ are the Cartan's connection
components defined as
\begin{eqnarray}
\Gamma^{*\kappa}_{\lambda\mu}(x,y)=\left(\gamma^{\kappa}_{\lambda\mu}-C^{\kappa}_{\lambda\rho}G^{\rho}_{\mu}-C^{\kappa}_{\rho\mu}G^{\rho}_{\lambda}+C_{\lambda\mu\rho}G^{\rho}_{\nu}g^{\nu\kappa}\right)(x,y)
\end{eqnarray}
and the $G^{\mu}_{\nu},G^{\mu}$ are
\begin{eqnarray}
G^{\mu}_{\nu}=& \frac{\partial G^{\mu}}{\partial y^{\nu}} \\
      2G^{\mu}=& \gamma^{\mu}_{\rho\sigma}y^{\rho}y^{\sigma}.
\end{eqnarray}

\begin{center}
\begin{table}
\begin{tabular}{|c|c|c|}\hline
$\Omega_M$ & $z_t/H_0$ & $\Omega_K=1-\Omega_M+z_t/H_0$ \\ \hline
0.110 & -0.9029492854 & -0.0129492854 \\ \hline
0.151 & -0.8443621881 &  0.0046378119 \\ \hline
0.232 & -0.7539951109 &  0.0140048891 \\ \hline
0.244 & -0.7710952485 & -0.0150952485 \\ \hline
0.251 & -0.7608626761 & -0.0118626761 \\ \hline
0.252 & -0.7327905974 &  0.0152094026 \\ \hline
0.253 & -0.7610339645 & -0.0140339645 \\ \hline
0.254 & -0.7556133186 & -0.0096133186 \\ \hline
0.255 & -0.7449999998 &  0.00000000002 \\ \hline
0.259 & -0.7313813921 &  0.096186079  \\ \hline
0.260 & -0.7400000040 &  -0.000000004  \\ \hline
0.261 & -0.7236653148 &  0.0153346852 \\ \hline
0.262 & -0.7378258680 &  0.0001741320 \\ \hline
0.263 & -0.7464096424 & -0.0094096424 \\ \hline
0.264 & -0.7249493659 &  0.0110506341 \\ \hline
0.2651 & -0.7441489179 & -0.0092489179 \\ \hline
0.270 &  -0.7425851580 & -0.012585158 \\ \hline
0.271 &  -0.7389289128 & -0.099289128 \\ \hline
0.272 &  -0.7279999954 &  0.0000000046 \\ \hline
0.273 &  -0.7337402645 & -0.0067402645 \\ \hline
0.274 &  -0.7135079061 &  0.0040405097 \\ \hline
0.275 &  -0.7151439555 &  0.098560445  \\ \hline
0.2761 & -0.7238999999 &  0.0000000001 \\ \hline
0.277 &  -0.7363820256 & -0.0133820256 \\ \hline
0.278 &  -0.7166084631 & 0.0053915369 \\ \hline
0.279 &  -0.7339876359 & -0.0129876359 \\ \hline
0.281 &  -0.7038915738 & 0.0151084262 \\ \hline
0.282 &  -0.7096732356 & 0.0083267644 \\ \hline
0.2831 & -0.7079203819 & 0.089796181 \\ \hline
0.284 &  -0.7285702833 & -0.0125702833 \\ \hline
0.285 &  -0.7232538088 & -0.083538088 \\ \hline
0.2861 & -0.7229838341 & -0.090838341 \\ \hline
0.289 &  -0.710999993  & 0.000000007 \\ \hline
\end{tabular}
\end{table}
\end{center}

\begin{center}
\begin{table}
\begin{tabular}{|c|c|c|}\hline
0.290 & -0.7032528013 & 0.0067471987 \\ \hline
0.291 & -0.7039021795 & 0.0050978205 \\ \hline
0.295 & -0.7002433383 & 0.0047566617 \\ \hline
0.2981 & -0.6924644204 & 0.0094355796 \\ \hline
0.325 & -0.6674671885 &  0.0075328115 \\ \hline
0.341 & -0.6560099554 &  0.0029900446  \\ \hline
0.345 & -0.6501638547 &  0.0048361453 \\ \hline
0.351 & -0.6617634475 & -0.0127634475 \\ \hline
0.36 & -0.6545617924  & -0.0145617924 \\ \hline
0.372 & -0.6385195052 & -0.0105195052  \\ \hline
0.38 &  -0.6199999969 &  0.0000000031  \\ \hline
0.392 & -0.6008829186 &  0.0071170814  \\ \hline
0.411 & -0.5985044824 & -0.0095044824  \\ \hline
0.421 & -0.5789999998 &  0.0000000002 \\ \hline
0.451 & -0.5563735170 & -0.0073735170 \\ \hline
0.512 & -0.4989395368 & -0.0109395368 \\ \hline
\end{tabular}
\caption{\small The $A-$test estimation for the parameter $z_t$.We insert the values of the $\Omega_M$
to the baryon acoustic oscillation peak $A=\sqrt{\Omega_M}\left[\frac{H_0^3d_L^2(z_1)}{H_1z_1^2(1+z_1)^2}\right]^{1/3}$
and calculate back the corresponding values of $z_t$, where $A=0.469\pm 0.017$\cite{Eisenstein}, $z_1=0.35$ the
typical luminous red galaxies redshift and $d_L(z)$ the luminosity distance defined in various
Relativity textbooks(e.g. see \cite{Carol}).The parameter $\Omega_K$  close to zero as expected
from the present observational predictions for a flat universe of WMAP \cite{tywmap}.}
\end{table}
\end{center}

\newpage

\begin{figure}
\begin{center}
\includegraphics[width=0.7 \textwidth, angle=0]{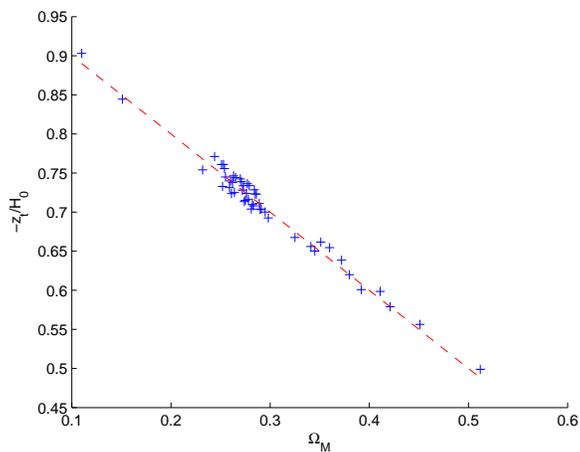}
\end{center}
\caption{\small The $z_t$ parameter versus $\Omega_M$;the $A$-test values of $-z_t/H_0$ are depicted by the ``plus sign'' line and
compared to the ``dashed line'' values coming from the $S$-test.Both tests almost reproduce  the
theoretical values of $z_t$ for a flat universe($\Omega_K=0$) calculated by the Friedman equation $-z_t/H_0=1-\Omega_M$.  )}
\end{figure}

\end{document}